# Unlocking with the Heart: Investigating User Perceptions of ECG-Based Biometric Authentication

Farnaz Farid*[1], Linet Momposhi[2], Tamara Watson[1], Alana Maurushat[1], Abubakar Bello[3], Farhad Ahamed[1]
[1] Cybersecurity and Behaviour, School of Social Sciences, Western Sydney University
[2] Graduate Cyber Security Analyst, Iglu Student Accommodation.
[3] School of Law & Criminal Justice, Edge Hill University

# Abstract

Password-based authentication remains challenging for many users, especially for those who face usability and accessibility issues. The burden of maintaining security can cause user fatigue, frustration and non-engagement. To address these issues, we conducted a pilot study to assess ECG-based biometric authentication as a passwordless alternative and to gather adult participants' perspectives on this approach. The research consisted of a survey study, an authentication prototype and a follow-up usability experience questionnaire. Our approach combined an online survey of 337 adults examining authentication practices and perceptions, with a separate usability evaluation in which 49 participants interacted with a simulated ECG-authentication and password interface in a health-care application prototype, then completed a post-test usability test questionnaire. The survey examined how adults across age groups use and perceive various authentication techniques, their experiences with biometrics, their views on security and convenience, and their willingness to adopt ECG authentication. In parallel, the usability prototype and questionnaire captured participants' responses after hands-on interaction with an ECG authentication prototype.

In the survey (study 1), 52.2% of participants indicated that they would be comfortable using ECG authentication. The acceptance was largely associated with prior use of biometric authentication for banking and with perceiving biometrics as more secure than passwords. although age did not play a decisive role. In the interactive prototype and questionnaire (study 2), participants generally perceived the simulated interaction as easy to learn and use and suitable for integration with mobile applications. These findings concern perceived usability and perceived security under controlled prototype conditions rather than operational performance, recognition accuracy, technical security, or resistance to attack.

The findings suggest that familiarity, transparency, privacy protection, understandable onboarding, and fallback authentication options may influence how users evaluate an ECG-authentication concept. Further research using operational ECG hardware, real biometric-template matching, larger and more age-balanced samples, and validated acceptance and usability instruments is required before conclusions can be drawn about real-world ECG-authentication performance or viability.

# 1. Introduction

Managing a large number of complicated passwords is a practical challenge for many users. This difficulty is particularly apparent in situations where people frequently need to access health, financial, and government services, as failed authentication can disrupt crucial activities. The present study examines ECG-based biometric authentication as a potential solution. Electrocardiogram is a record of the heart's electrical voltage fluctuations. ECG- based biometric authentication uses a person's finger to read heart signals as the process of verifying a user's identity. As heart signals are unique to every individual, it is a promising way of making sure a person is who they claim to be.

Instead of viewing ECG authentication merely as a technical issue, we focus on how adults across age groups understand, assess, and experience the method compared with the authentication methods they are accustomed to. To investigate the perceived usability of ECG biometric authentication and the factors influencing people's willingness to adopt it, we conducted a pilot study using a prototype of a healthcare application.

This study is an extension of our previous research into multimodal biometric authentication for use in personalised healthcare (Farid and Ahamed, 2019; Ahamed et al., 2022). In that earlier work, ECG and PPG signals were suggested as possible authentication methods, but it had not been examined how potential users would respond to them. The current study fills this gap by combining survey findings with usability data from a prototype. By doing this, it goes beyond the question of whether ECG authentication is technically feasible and instead considers whether users find it understandable, acceptable, and usable in a healthcare applications context.

- RQ1: How do adults across different age groups experience and perceive current authentication methods?
- RQ2: Do older adults differ from younger adults in their experience and perceptions of current authentication methods?
- RQ3: What factors influence acceptance of ECG-based authentication as a passwordless authentication method?
- RQ4: Do older adults differ from younger adults in their acceptance of ECG authentication, biometric experience, and perceptions of security?
- RQ5: How usable is ECG-based authentication following prototype evaluation?

We have conducted this research in four stages.

- **Stage 1:** We administered a Survey (Study 1) across adult digital-service users to examine authentication practices, biometric familiarity, and attitudes toward ECG-based authentication
- **Stage 2:** We designed and developed a healthcare app prototype incorporating password- and ECG-based authentication pathways.
- **Stage 3:** We conducted a prototype usability evaluation (Study 2) with adult participants across age groups.
- **Stage 4:** Participants completed a post-test usability questionnaire (Study 2), including System Usability Scale-informed items and additional measures of perceived security, learnability, efficiency, confidence, and intention to use.

The contributions of this study are as follows:

- First, it provides survey-based evidence of adult users' perceptions of ECG-based authentication through a sample of 337 participants, examining authentication behaviours, biometric familiarity, and willingness to adopt ECG authentication.
- Second, the study presents findings from a prototype-based usability evaluation involving a simulated ECG-authentication interface, offering insights into participants' perceptions of usability, perceived security, and acceptance of an ECG-authentication workflow.
- Third, the study identifies prior biometric experience and perceived biometric security as factors associated with stated willingness to adopt ECG authentication.
- Fourth, the study provides exploratory evidence regarding ECG-authentication perceptions across age groups. Although older participants reported lower levels of prior biometric use, age was not significantly associated with ECG-authentication acceptance or perceived biometric security in Study 1. These findings should be interpreted cautiously and require confirmation in larger and more balanced samples.

# 2. Background

Authentication and password practices are essential components of usable security research. According to the NordPass survey, the average person currently manages around 70 to 80 passwords (Higgins, 2020). Despite recommendations to use unique passwords for each account, 65% of users reuse passwords across multiple sites (Google, 2022). Strong passwords are also recommended to reduce cyber vulnerabilities (Ray et al., 2021). However, password creation and use involve a persistent security-usability trade-off: users must balance complexity, memorability, frequency of use, and perceived account importance when choosing passwords (Wash and Rader, 2021).

Some studies show that password complexity decreases with age (Juozapavičius et al., 2022), which may be related to cognitive decline, security-notification fatigue and authentication burden (Abdrabou et al., 2022; Wash and Rader, 2021). As they age, people may become more concerned about security, privacy, and safety while engaging in online activities (Ray et al., 2021), and many experience technology anxiety (Caspi et al., 2019). Yet cybersecurity behaviour among older adults remains under-researched (Morrison et al., 2021). Various challenges hinder older populations from engaging in online activities (Gell et al., 2015), and many technologies are not designed with older adults' accessibility and usability needs in mind (Ray et al., 2021; Wang et al., 2024).

Alt and Schneegass (2022) observed that an alternative way to address password memorisation and fatigue is through passwordless authentication techniques such as biometrics. Biometric authentication authenticates a person by proving "who someone is" through a unique physical or physiological feature such as a fingerprint, face, iris, or ECG signal. However, biometric methods have limitations related to data collection, inclusivity, accessibility, and security concerns (Ingale et al., 2020; Wang et al., 2024). For example, fingerprint deterioration and challenges in iris recognition in later life can create age-related exclusion risks (Galbally et al., 2018; Azimi et al., 2021). Hence, beyond conventional physical features, physiological signals such as electrocardiogram (ECG), electroencephalogram (EEG), and photoplethysmogram (PPG) have attracted attention from the research community. ECG biometric authentication has been examined as a promising modality because ECG signals are individualised, difficult to counterfeit, and relevant to emerging healthcare and wearable contexts. However, real-world deployment still faces challenges related to datasets, signal quality, privacy, and usability (Ingale et al., 2020). One primary reason for this trend is the increased use of Medical Internet of Things (IoT) devices (Farid and Ahamed, 2019; Zhang et al., 2021).

The increased older adult population and growing dependence on digital technologies have established a need to study older adults' attitudes toward technology and authentication methods (Grindrod et al., 2018; Das et al., 2019; Halmdienst et al., 2019; Lim et al., 2022; Bailey et al., 2021; Wang et al., 2024). However, few studies have examined older adults' attitudes toward IoT-based healthcare apps with biometric authentication options (Kružíková et al., 2022; Grindrod et al., 2018). Recent studies in Canada, the US, the UK, and Singapore have analysed the challenges and experiences of older populations with password authentication technologies and mobile health applications (Grindrod et al., 2018; Das et al., 2019; Lim et al., 2022; Bailey et al., 2021; Murabito et al., 2024). To our knowledge, limited research has explored age-related differences in perceptions of ECG-based biometric authentication. Accordingly, this study examines Australian adults' mental models of authentication technologies and provides insights into authentication behaviours, biometric familiarity, perceived security, and perceptions of ECG-based biometric authentication.

We apply the Technology Acceptance Model (TAM), which explains technology adoption in terms of users' perceived usefulness and perceived ease of use (Davis, 1989). In the context of biometric and passwordless authentication, the core constructs of TAM require extension, as acceptance of a technology is also shaped by perceived security, trust, perceived risk, privacy concerns, prior technology experience, and the practical context in which authentication is used. Recent biometric-authentication studies have therefore extended TAM to include constructs such as trust, perceived security, perceived risk, and personal innovativeness when examining user acceptance of facial, palm-vein, and other biometric technologies (James et al., 2008; Nakisa et al., 2023; Rukhiran et al., 2023).

For ECG-based authentication, an extended TAM is more appropriate, as ECG authentication is both a security technology and a physiological biometric linked to health-related data. Users may evaluate it not only by whether it is easy to use or reduces password burden, but also by whether it appears trustworthy, privacy-preserving, reliable, understandable, and compatible with familiar devices. Accordingly, in this paper, we apply an extended TAM-informed framing to interpret acceptance of ECG authentication in terms of perceived usefulness, perceived ease of use, perceived security and trust, privacy and risk concerns, and prior biometric familiarity.

Unlike fingerprint and facial recognition, ECG authentication relies on a physiological signal that may be integrated into healthcare ecosystems, wearable technologies, and medical IoT applications. While ECG authentication has attracted attention because of its potential resistance to conventional spoofing attacks and its relevance to healthcare settings, its adoption ultimately depends on whether users perceive it as understandable, trustworthy, and practical. Furthermore, because ECG data may be viewed as health-related information, concerns about transparency, data governance, and privacy may influence acceptance differently than with more established biometric methods. Understanding these perceptions is therefore critical to designing future passwordless healthcare systems.

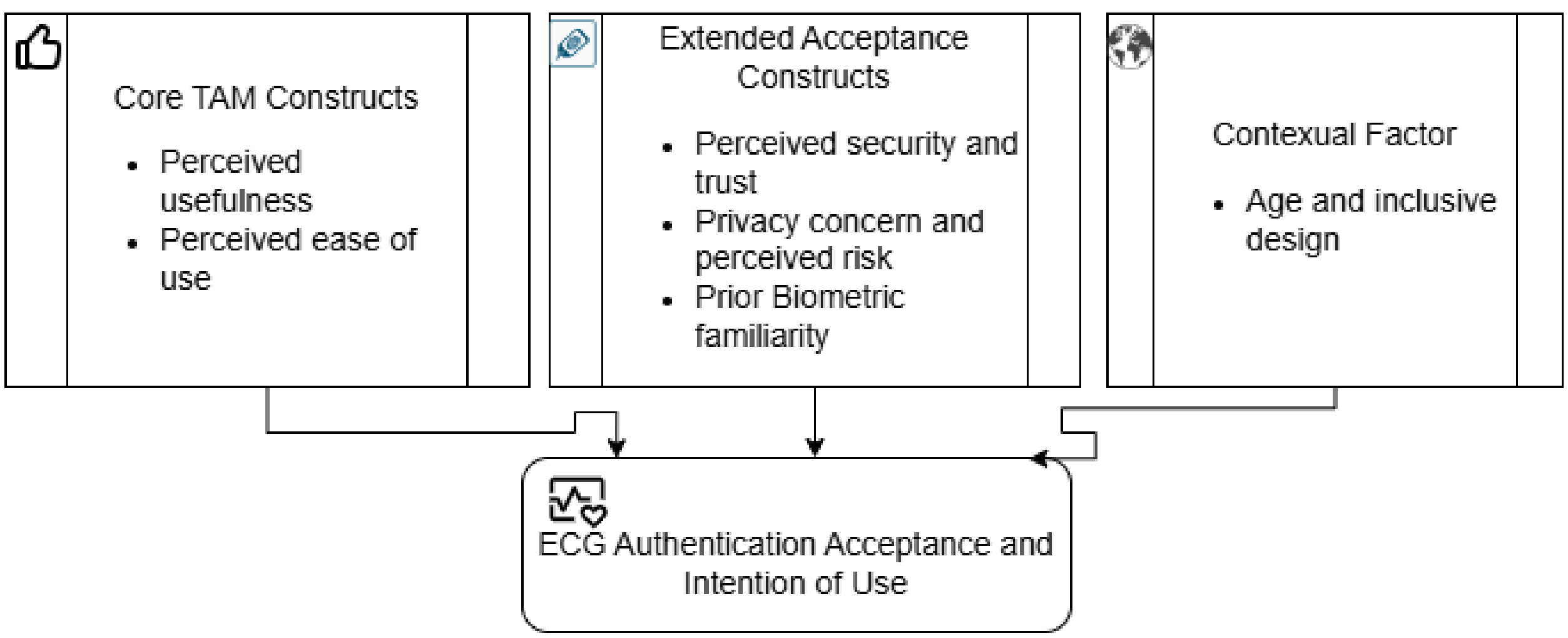

***Figure 1. Extended TAM-informed conceptual model for interpreting ECG authentication acceptance.***

# 3. Stage 1: Survey Study 1

## 3.1 Study Design

We applied a quantitative pilot study incorporating supplementary open-text responses. In the first phase, we conducted an anonymous online survey of 337 adults that captured demographic information, current authentication practices, biometric use, perceived security, perceived convenience, and comfort with ECG authentication.

## 3.2 Participants and Recruitment

A total of 348 participants took part in the survey using Qualtrics. Initial data screening removed 10 incomplete responses where the survey was not finished. Of the remaining 338 responses, 1 respondents indicated they were under 18 years of age and were automatically excluded from the final analysis. Consequently, the final analysed sample comprised 337 valid responses. We used both real-world and virtual methods to recruit them. We advertised on social media by briefly explaining our research, providing an email address for further information, and including a link to the online study on Qualtrics. Participants completed a 20-minute survey on their authentication practices and preferences. The questionnaire is given in the appendix. We recruited participants via an open call for individuals aged 18 and over. We divided participants into seven broad age groups based on the demographic data collected. Table 1 displays the distribution of participants across key demographic characteristics.

*Table 1. Participant distribution across demographic characteristics in the online survey (N = 337).*

| Characteristic | Category | Total count | Percentage |
|---|---|---|---|
| **Age group** | 18–24 | 50 | 14.8 |
| | 25–34 | 67 | 19.8 |
| | 35–44 | 59 | 17.5 |
| | 45–54 | 56 | 16.6 |
| | 55–64 | 55 | 16.3 |
| | 65–74 | 36 | 10.7 |
| | 75+ | 12 | 3.6 |
| | Missing | 2 | 0.6 |
| **Education** | Bachelor's degree | 122 | 36.2 |
| | High school diploma | 90 | 26.6 |
| | Master's degree | 62 | 18.3 |
| | Some high school | 20 | 5.9 |
| | PhD or higher | 19 | 5.6 |
| | Trade school/vocational | 18 | 5.3 |
| | Apprenticeship | 4 | 1.2 |
| | Missing | 2 | 0.6 |
| **Employment status** | Employed for wages | 177 | 52.5 |
| | Homemaker | 7 | 2.1 |
| | Retired | 43 | 12.7 |
| | Student | 40 | 11.9 |
| | Self-employed | 37 | 10.9 |
| | Out of work and looking for work | 21 | 6.2 |
| | Unable to work | 5 | 1.5 |
| | Out of work, not looking for work | 3 | 0.9 |
| | Military | 1 | 0.3 |
| | Missing | 3 | 0.9 |
| **Location** | Not in Western Sydney | 252 | 74.8 |
| | Western Sydney | 82 | 24.3 |
| | Missing | 3 | 0.9 |

The sample surveyed (N = 337) included people from a wide variety of age groups, with the largest proportion in the 25 to 34-year age group (19.8%). The largest educational group comprised participants with a bachelor's degree (36.2%) and were employed on a wage basis (52.5%). Most of the respondents lived outside Western Sydney (74.8%).

## 3.3 Survey Instrument

The demographic section of the online survey collected age group, education level, employment status, and residential location. The study included age group because one aim was to examine whether authentication experiences and ECG acceptance differed across adult age groups. The authentication-behaviour section asked participants about their current use of passwords, biometrics, password managers, and multifactor authentication. It also captured password reuse, difficulties managing multiple credentials, and perceived problems with existing authentication methods. The biometric-authentication section explored the application of biometric authentication in everyday digital services. Participants reported whether they currently used biometric authentication, what type they use, and which services they use it for.

The ECG-authentication section focused on participants' awareness of ECG-based authentication, willingness to use it, preferred implementation methods, potential use contexts, and perceived barriers to adoption. We asked participants reluctant to use ECG authentication to identify their concerns. Several attitudinal items measured technology confidence, perceived security of authentication methods, and perceptions of convenience and usability. Most attitudinal items used five-point Likert scales ranging from Strongly Disagree to Strongly Agree.

For inferential analyses, ECG authentication acceptance served as the primary outcome variable. Predictor variables included demographic characteristics, biometric adoption, awareness of ECG authentication, technology confidence, password-management behaviours, and security-related attitudes. Additional analyses compared younger adults (<55 years) and older adults (≥55 years) to examine age-related differences in biometric experience, ECG acceptance, and perceptions of security.

## 3.4 Data Analysis

We treated ECG authentication acceptance as a binary outcome based on whether participants indicated they would feel comfortable using ECG-based authentication. We first used descriptive statistics to summarise participant characteristics and response patterns. We then conducted Pearson chi-square tests of independence without continuity correction to examine associations between ECG acceptance and selected demographic variables, security behaviours, prior biometric experience, and perceptions of ECG authentication.

All quantitative analyses were conducted using Jamovi Cloud version 28.8. Effect sizes were reported using Phi ($\varphi$) for 2 × 2 contingency tables and Cramer's V for larger tables. Following Cohen's conventions, we interpreted values of approximately .10, .30, and .50 as small, medium, and large effects, respectively. Because one objective of the study was to explore ECG authentication as a passwordless option for older adults, we also conducted an age-based analysis. We classified participants aged 55 years and older as older adults, and those younger than 55 years as younger adults. We selected the 55-year threshold to support the study's exploratory focus on later-adulthood authentication experiences while retaining sufficient observations in the older group. This dichotomisation was analytical and should not be interpreted as a universal definition of older adulthood.

Finally, we fitted a binary logistic regression model to identify independent predictors of acceptance of ECG authentication. We entered variables that showed significant associations in the chi-square analyses into the model. We calculated odds ratios (ORs) and 95% confidence intervals (CIs) and assessed statistical significance at $\alpha = .05$. This predictor-selection approach was used to reduce model complexity in an exploratory pilot study. However, selecting predictors based on preliminary bivariate significance tests may exclude conceptually relevant variables and introduce model-selection bias. Consequently, the regression findings should be interpreted as exploratory and confirmed in future studies using larger samples and theory-driven modeling approaches.

We interpreted the quantitative findings through an extended TAM-informed analytical framework. We examined ECG authentication acceptance in relation to constructs corresponding to perceived usefulness, perceived ease of use, perceived security and trust, privacy or perceived risk, and prior biometric familiarity. Because the survey did not use a full set of validated TAM measurement scales and the study was exploratory, we used the extended TAM framework to organise interpretation of the findings rather than to validate a complete acceptance model.

## 3.6 Findings

### 3.6.1 Current Authentication Practices

The findings indicate that authentication behaviour remains mixed. Among respondents who answered the password reuse question, 197 of 336 respondents (58.6%) reported reusing passwords across multiple services, while 139 (41.4%) did not. The most common authentication combination was PIN, password, and biometric authentication (98 responses), followed by PIN and password together (73 responses), password only (43 responses), and PIN, password, pattern lock, and biometric authentication together (25 responses). These findings suggest that participants commonly rely on layered authentication approaches that combine multiple authentication mechanisms. However, the high prevalence of password reuse indicates that password-related security risks remain substantial despite the increasing use of biometric authentication.

### 3.6.2 Perceptions of Biometrics

Many respondents were familiar with biometric authentication. Among the 298 participants who reported using biometrics and answered the question on biometric modality, fingerprint recognition was the most commonly reported biometric method (109 participants), followed by facial recognition (81 participants). A further 72 participants reported using both fingerprint and facial recognition. These findings indicate that fingerprint- and face-based biometrics were widely used authentication methods within the sample. Most participants also perceived biometric authentication to be more secure than traditional password-based authentication, with 240 of 337 valid respondents (71.0%) indicating that biometric methods were either somewhat or much more secure than passwords.

### 3.6.3 Acceptance of ECG Authentication

A total of 337 valid responses were available for analysis. Of these, 176 participants (52.2%) reported they would be comfortable using ECG authentication, while 161 (47.8%) indicated they would not.

### 3.6.4 Barriers to ECG Adoption

Participants identified factors that would prevent them from using ECG authentication. The most frequently reported single barriers were 'I have no idea how it works' (n = 51), 'I am not sure' (n = 46), 'Other' (n = 39), 'It will add unnecessary complexity' (n = 32), and 'I don't feel comfortable saving my biometric information' (n = 19). Concerns about biometric information being hacked also appeared in 25 responses in both standalone and combined responses. These findings suggest that reluctance to adopt ECG authentication is influenced primarily by limited understanding of the technology, uncertainty, perceived complexity, and privacy and security concerns rather than by technical performance considerations alone. These are illustrated in Table 5.

*Table 2. Summary of ECG authentication acceptance and adoption barriers.*

| Construct | Finding | Implication |
|---|---|---|
| **ECG acceptance** | 176/337 (52.2%) accepted ECG authentication; 161/337 (47.8%) did not. | Acceptance was moderate, indicating mixed attitudes towards ECG authentication. |
| **Non-adoption intent** | "I will not use them" was the most common response in the ECG use-context question (n = 129). | A substantial proportion of participants remain unconvinced about ECG authentication. |
| **Knowledge barrier** | "I have no idea how it works" was the most frequently reported barrier (n = 51). | User education and awareness may improve adoption. |
| **Complexity barrier** | 32 participants reported unnecessary complexity. | Simpler onboarding and interaction design may improve acceptance. |
| **Privacy and security barrier** | Multiple participants reported concerns about storing biometric data (n = 19) and the risk of hacking (n = 25). | Emphasize privacy safeguards and transparency. |

### 3.6.5 Chi-Square Results

To identify factors associated with ECG authentication acceptance, we conducted a series of chi-square tests of independence between ECG acceptance and demographic characteristics, security behaviours, biometric familiarity, and security perceptions. Effect sizes were reported using Phi (φ) for 2 × 2 contingency tables and Cramer's V for larger tables. Table 6 presents the chi-square analysis of predictors of ECG-authentication acceptance.

Perceived biometric security was significantly associated with ECG-authentication acceptance, $\chi^2(5, N = 337) = 17.44$, $p = .004$, Cramer's V = .227. Participants who perceived biometric authentication as more secure than traditional password-based authentication were more likely to report willingness to adopt ECG authentication. Specifically, 240 of 337 valid respondents (71.0%) considered biometric authentication to be somewhat or much more secure than passwords.

Similarly, prior experience with biometric authentication through banking services was significantly associated with ECG acceptance, $\chi^2(1) = 9.94$, $p = .002$, $\varphi = .172$. Participants who had already used biometric authentication for banking were more likely to say they would be comfortable using ECG authentication.

No significant associations were identified between ECG-authentication acceptance and age group, education level, employment status, residential location, password-reuse behaviour, password-manager awareness, password-manager use, or multifactor-authentication use (all $p > .05$). Password reuse was not significantly associated with ECG-authentication acceptance, $\chi^2(1, N = 336) = 0.95$, $p = .330$, $\varphi = .053$. Similarly, education, employment status, residential location, password-manager awareness, and password-manager use were not significantly associated with ECG-authentication acceptance. Table 6 examines predictors of ECG-authentication acceptance, whereas the age-based analyses reported later examine associations between age group and selected authentication behaviours and perceptions. These analyses address different research questions and therefore yield different chi-square statistics.

*Table 3. Chi-square analysis of predictors of ECG authentication acceptance.*

| Variable | N | χ² | Df | p | Effect size |
|---|---|---|---|---|---|
| Perceived biometric security (Study 1–Q22) | 337 | 17.44 | 5 | .004 | V = .227 |
| Biometric banking use (Q19) | 337 | 9.94 | 1 | .002 | φ = .172 |
| Age group (<55 vs ≥55) | 335 | 2.10 | 1 | .147 | φ = .079 |
| Education | 335 | 7.16 | 6 | .306 | V = .146 |
| Employment status | 334 | 9.35 | 8 | .314 | V = .167 |
| Residential location | 334 | 2.17 | 1 | .140 | V = .081 |
| Password reuse | 336 | 0.95 | 1 | .330 | φ = .053 |
| Password manager awareness | 337 | 0.07 | 1 | .785 | φ = .015 |
| Password manager use | 333 | < 0.001 | 1 | .985 | φ = .001 |

### 3.6.6 Logistic Regression Analysis

A binary logistic regression analysis was conducted to identify independent predictors of ECG-authentication acceptance. Only variables that showed significant associations in the chi-square analyses were included in the final model. These variables were biometric-authentication use for banking services (Q19) and perceived biometric security relative to passwords (Q22).

The dependent variable was ECG-authentication acceptance, measured by responses to the question “If in use, will you be comfortable with ECG authentication?” (Yes = 1, No = 0). For the regression analysis, perceived biometric security (Q22) was recorded into a binary variable. Participants who considered biometric authentication somewhat more secure or much more secure than passwords were coded as 1, while all other responses (much less secure, somewhat less secure, not more or less secure, and I don't know) were coded as 0. The chi-square analysis retained all six original Q22 categories. Table 7 illustrates the results.

*Table 4. Binary logistic regression predicting ECG authentication acceptance.*

| **Predictor** | **β** | **OR** | **95% CI** | **p** |
|---|---|---|---|---|
| Biometric banking use (Q19) | 0.610 | 1.84 | 1.18–2.87 | .007 |
| perceived biometric security (Study 1–Q22) | 0.604 | 1.83 | 1.12–2.98 | .015 |
| Constant | -0.627 | 0.53 | 0.34–0.83 | .005 |

Model statistics:

- N = 337
- Likelihood Ratio $\chi^2 \approx 15.94$

- $p < .001$
- McFadden $R^2 = .034$

Participants who used biometric authentication for banking had 1.84 times the odds of accepting ECG authentication compared with non-users (OR = 1.84, 95% CI [1.18, 2.87], $p = .007$). Similarly, participants who perceived biometric authentication as somewhat or much more secure than passwords had 1.83 times the odds of accepting ECG authentication compared with participants who selected other security-perception responses (OR = 1.83, 95% CI [1.12, 2.98], $p = .015$).

Demographic variables, including age, education, employment status, and residential location, along with other nonsignificant bivariate predictors, were excluded from the final regression model because they showed no significant associations with ECG-authentication acceptance in the preliminary chi-square analyses. Likewise, password-reuse behaviour, password-manager awareness, password-manager use, and multifactor-authentication use did not demonstrate significant bivariate associations with ECG-authentication acceptance and were therefore excluded from the regression analysis.

These findings suggest that acceptance of ECG authentication was more strongly associated with prior biometric-banking experience and perceived biometric security than with the demographic and cybersecurity-practice variables examined in this study. Because predictor selection was based on statistical significance in preliminary bivariate analyses, the regression findings should be interpreted as exploratory.

### 3.6.7 Age-Based Analysis

To explore the suitability of ECG authentication as a passwordless authentication mechanism for older adults, we grouped participants into younger adults (18–54 years) and older adults (55 years and above). We examined age-related differences in ECG acceptance, password reuse behaviour, familiarity with biometric banking, and perceptions of biometric security. Tables 5–8 examine age-related differences in authentication behaviours and perceptions.

### ECG acceptance by age

*Table 5. ECG authentication acceptance by age group.*

| Age group | Accept ECG | Reject ECG | Acceptance rate |
|---|---|---|---|
| **Younger (<55 years)** | 128 | 104 | 55.2% |
| **Older (≥55 years)** | 48 | 55 | 46.6% |

ECG-authentication acceptance did not differ significantly between younger and older adults, $\chi^2(1, N = 335) = 2.10$, $p = .147$, $\varphi = .079$. Although younger adults reported a slightly higher acceptance rate (55.2%) than older adults (46.6%), the association was not statistically significant, and the effect size was small.

### Password reuse by age

*Table 6. Password reuse behaviour by age group.*

| Age group | Reuse passwords | Do not reuse |
|---|---|---|
| **Younger (<55 years)** | 145 (62.8%) | 86 (37.2%) |
| **Older (≥55 years)** | 52 (50.5%) | 51 (49.5%) |

Password-reuse behaviour differed significantly between younger and older adults, $\chi^2(1, N = 334) = 4.44$, $p = .035$, $\varphi = .115$. Younger adults were more likely than older adults to report password reuse. Although the effect size was small, the finding suggests that older adults reported somewhat stronger password-management practices than younger participants.

## Biometric banking familiarity by age

*Table 7. Biometric authentication use for banking by age group.*

| Age group | Use biometrics for banking | Do not use |
|---|---|---|
| **Younger (<55 years)** | 136 (58.6%) | 96 (41.4%) |
| **Older (≥55 years)** | 24 (23.3%) | 79 (76.7%) |

Younger adults were substantially more likely than older adults to report biometric-authentication use for banking services, $\chi^2(1, N = 335) = 35.7$, $p < .001$, $\varphi = .326$. This medium effect suggests that younger participants had substantially greater exposure to biometric technologies in everyday settings.

## Perceived security of biometrics by age

The selected age grouping was not significantly associated with whether participants perceived biometric authentication as somewhat or much more secure than passwords, $\chi^2(1, N = 335) = 0.56$, $p = .455$, $\varphi = .041$. Participants in both age groups generally perceived biometric authentication as more secure than password-based authentication.

*Table 8. Perceived security of biometric authentication by age group.*

| Age group | Biometrics more secure | Other responses |
|---|---|---|
| **Younger (<55 years)** | 167 (72.0%) | 65 (28.0%) |
| **Older (≥55 years)** | 70 (68.0%) | 33 (32%) |

## 3.7 Extended TAM interpretation of ECG authentication acceptance

The findings can be interpreted using the extended Technology Acceptance Model (TAM) introduced earlier. Consistent with the model, participants appeared to evaluate ECG authentication through the lenses of perceived usefulness and perceived ease of use. Following interaction with the prototype, participants generally viewed the simulated ECG-authentication interaction as easy to learn and simple to use, and many expressed willingness to use ECG authentication in place of traditional passwords. These findings suggest that participants recognised ECG authentication's potential to reduce password burden while maintaining a relatively straightforward authentication experience.

Beyond usefulness and ease of use, perceived security emerged as an important factor influencing acceptance. Participants frequently perceived ECG authentication as more secure than password-based authentication, and perceptions of biometric security were significantly associated with willingness to adopt ECG authentication. This finding aligns with previous biometric-authentication research suggesting that trust and security perceptions play a central role in user acceptance of emerging authentication technologies.

The study also highlights privacy concerns and perceived risk. Although participants recognised potential security benefits, many respondents in study 1 expressed concerns about storing biometric information, uncertainty about how ECG authentication works, and the possibility of biometric data being compromised. These concerns indicate that acceptance of ECG authentication depends not only on security performance but also on transparency, user control, and confidence in data-handling practices.

Prior familiarity with biometrics emerged as another important factor. In Study 1, participants with previous experience using biometric technologies, particularly biometric banking authentication, were significantly more likely to accept ECG authentication. This finding suggests that familiarity with biometric systems may reduce uncertainty and increase trust when users encounter novel forms of biometric authentication.

# 4. Stage 2: ECG Authentication Prototype

We developed a healthcare app prototype that allowed users to register and sign in using both ECG-based and password-based authentication options. The prototype interfaces are shown in Figures 2 and 3.

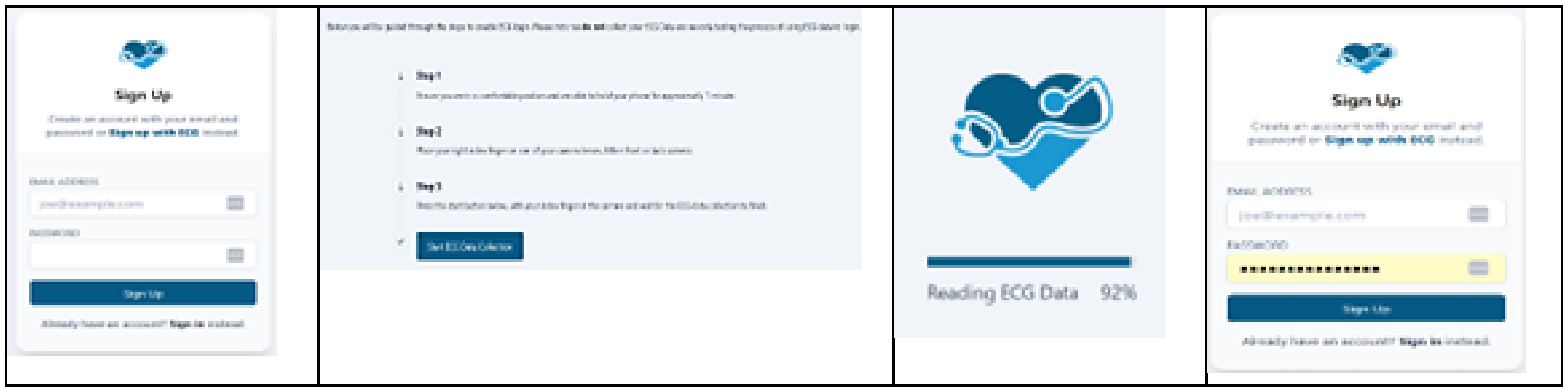

*Figure 2. ECG sign-up interface for the health app prototype.*

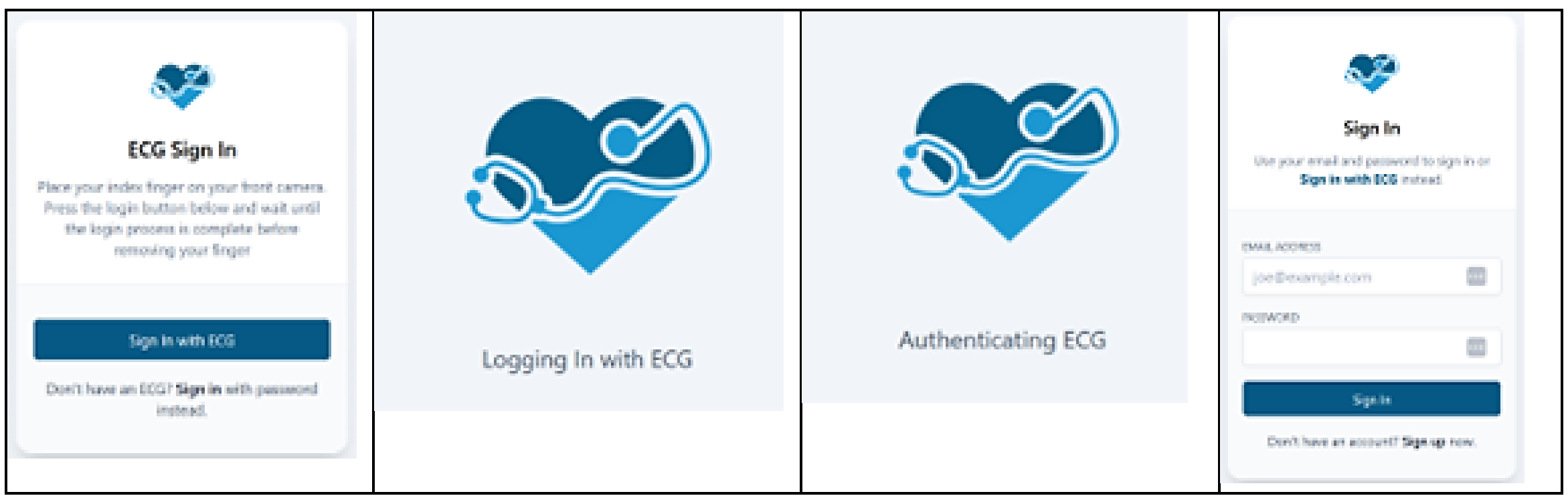

*Figure 3. ECG sign-in interface for the health app prototype.*

# 5. Stage 3: Prototype Usability Evaluation

A prototype usability test was performed in person with 49 participants who used both ECG and traditional passwords to authenticate. A questionnaire about user experience was provided after the prototype test which will be examined in Section 6. The test used an ECG healthcare application with password login options. Table 9 summarises the demographic distribution for the prototype usability survey.

*Table 9. Participant distribution across age groups in the prototype usability survey.*

| Age group | Count |
|---|---|
| **18–24** | 10 |
| **25–34** | 23 |
| **35–44** | 8 |
| **45–54** | 3 |
| **55–64** | 2 |
| **65–74** | 3 |
| **Total** | 49 |

A total of 52 individuals initiated the usability test. We excluded three incomplete submissions (15%, 22%, and 44% completion), leaving 49 completed responses for analysis. Because several respondents did not complete all items, item-level sample sizes ranged from 47 to 49. We conducted analyses requiring complete responses across all 15 usability items using 45 complete cases.

The prototype usability test was conducted in person. A researcher was present throughout the testing session to provide instructions and ensure that participants completed the tasks in the intended sequence, although participants were instructed to complete all authentication tasks independently and without assistance. Participants used a mobile device and a laptop computer to interact with the ECG health app prototype and complete the study activities.

The authentication tasks were completed in the following order:

1. Register using the participant identifier.
2. Create a password meeting the stated password requirements of at least 8 characters, including a letter, at least 1 number, and a special character.
3. Complete the simulated ECG setup task by placing a finger over the device's camera while the prototype displayed the ECG setup process (note: no ECG was taken).
4. Log in using ECG authentication.
5. Log out.
6. Log in using password authentication.
7. Log out.
8. Complete the post-test questionnaire that was focused on the authentication experience rather than the overall design of the application.

## 5.1 Measures

We integrated PostHog with the prototype to capture session-level interaction data and record task performance measures. Recorded variables included time taken for password registration, ECG setup, password login, and instruction review. Additionally, we recorded password creation and login attempts.

Password-creation analyses were based on the 45 participants with complete password-creation data. Twenty-seven participants created a strong password on their first attempt, while 18 required more than one attempt. These observations suggest that the password pathway introduced both time cost and procedural friction. However, comparisons with ECG setup should be interpreted cautiously because the ECG task was represented as a highly standardised prototype interaction rather than a fully operational ECG-authentication process. Table 10 illustrates the results.

*Table 10 Prototype usability evaluation summary.*

| **Measure** | **Reported value** | **Interpretation** |
|---|---|---|
| Password setup time | Mean = 64.36 seconds | Password creation took longer than the recorded ECG setup task. |
| ECG setup time | 30 seconds | Recorded as a standardised prototype interaction; |
| Password login time | Mean = 22.30 seconds | Password login was faster than password setup but still added interaction time. |
| Instruction time | Mean = 22.95 seconds | Participants needed additional time to read or follow setup instructions. |
| Strong-password creation | 27 first-attempt successes; 18 required more than one attempt | Multiple attempts indicate password-strength friction for a substantial subset of users. |

# 6. Stage 4: Post-Prototype Usability Experience Questionnaire

Participants from stage 3 were invited to participate in a user experience questionnaire.

## 6.1 Design

The post-test questionnaire included two perceived-security items and a 15-item SUS-informed usability instrument adapted for the ECG-authentication prototype. Responses were recorded on a five-point Likert scale ranging from 1 (Strongly Disagree) to 5 (Strongly Agree). Ten items were negatively worded and were reverse-scored before scale construction. The inversion of negatively worded questions helps to reduce acquiescence bias, the tendency of respondents to agree with every statement regardless of content. The instrument was informed by the System Usability Scale (SUS) (Brooke, 1996) but extended beyond the standard 10-item SUS to address ECG-authentication-specific usability, learnability, confidence, integration with mobile applications, and perceived burden. The 15 items assessed perceived effort, annoyance, fatigue, necessity, complexity, technical-support requirements, consistency, learnability, ease of use, confidence, time demands, intention to use, and integration with mobile applications. Consequently, the resulting 0–100 usability index should be interpreted as a transformed SUS-informed usability measure rather than a conventional SUS score. Open-text responses are presented as illustrative feedback and were not subjected to formal qualitative coding or thematic analysis. Table 11 shows the *SUS-informed ECG usability instrument.*

*Table 11. SUS-informed ECG usability instrument.*

| Item | Description | Reverse scored |
|---|---|---|
| Q12 | ECG authentication takes a long time to use | Yes |
| Q13 | ECG authentication is annoying to use | Yes. |
| Q14 | ECG authentication is tiring to use | Yes. |
| Q15 | I would like to use ECG authentication regularly | No |
| Q16 | ECG authentication is unnecessary | Yes |
| Q17 | ECG authentication is complex | Yes |
| Q18 | ECG authentication is easy to use | No |
| Q19 | I would need technical support to use ECG authentication | Yes |
| Q20 | ECG authentication is inconsistent | Yes |
| Q21 | Navigation and setup are inconsistent | Yes |
| Q22 | ECG authentication is easy to learn | No |
| Q23 | ECG authentication is time-consuming | Yes |
| Q24 | I feel confident using ECG authentication | No |
| Q25 | I would need substantial learning before using ECG authentication | Yes |
| Q26 | ECG authentication integrates well with mobile applications | No |

## 6.2 Data Analysis

The ECG pathway represented a standardised prototype interaction rather than a fully operational ECG recognition system. Accordingly, the findings concern perceived usability and perceived security, not recognition accuracy, signal stability, attack resistance, or real-world authentication performance.

### 6.2.1 ECG usability analysis

Descriptive statistics were calculated for all usability items using responses from the completed post-test survey. Responses were recorded on a five-point Likert scale ranging from 1 (Strongly Disagree) to 5 (Strongly Agree). Because several participants did not answer every usability item, the valid sample size ranged from 47 to 49 responses per item. Analyses requiring complete responses across all 15 usability items used 45 complete cases. Table 12 illustrates the results.

*Table 12. Means and standard deviations for ECG usability items.*

| **Item** | Valid N | **Mean** | **SD** |
|---|---|---|---|
| ECG authentication takes a long time to use (Q12) | 49 | 2.90 | 1.18 |
| ECG authentication is annoying to use (Q13) | 47 | 3.45 | 1.06 |
| ECG authentication is tiring to use (Q14) | 49 | 3.51 | 1.06 |
| I would like to use ECG | 49 | 3.88 | 0.86 |
| authentication regularly (Q15) | | | |
| ECG authentication is unnecessary (Q16) | 49 | 3.35 | 1.13 |
| ECG authentication is complex (Q17) | 49 | 3.18 | 1.18 |
| ECG authentication is easy to use (Q18) | 49 | 3.94 | 0.85 |
| I would need technical support to use ECG authentication (Q19) | 49 | 3.39 | 1.17 |
| ECG authentication is inconsistent(Q20) | 49 | 3.29 | 1.14 |
| Navigation and setup are inconsistent(Q21) | 49 | 3.35 | 1.20 |
| ECG authentication is easy to learn (Study 2–Q22) | 48 | 3.94 | 0.86 |
| ECG authentication is time-consuming (Q23) | 49 | 3.31 | 1.12 |
| I feel confident using ECG authentication (Q24) | 49 | 3.90 | 0.90 |
| I would need substantial learning before using | 48 | 2.96 | 1.22 |
| ECG authentication (Q25) | | | |
| ECG authentication integrates well with mobile applications(Q26) | 49 | 3.82 | 0.67 |

Note. Item means are presented using the original response coding. Negatively worded items were reverse-scored only for calculation of the overall usability index; item-level means and standard deviations are reported in their raw form.

Participants reported the strongest agreement for ease of use (Q18: M = 3.94, SD = 0.85), learnability (Study 2–Q22: M = 3.94, SD = 0.86), confidence in using ECG authentication (Q24: M = 3.90, SD = 0.90), and integration with mobile applications (Q26: M = 3.82, SD = 0.67). Negatively worded items, including perceived complexity (Q17), technical-support requirements (Q19), inconsistency (Q20–Q21), and time demands (Q23), were reverse-scored before calculating overall usability measures so that higher values consistently represented more favourable usability perceptions. The resulting item-level pattern indicated generally favourable perceptions of the simulated ECG-authentication prototype, although individual responses varied.

After reverse-scoring negatively worded items, the ECG usability scale achieved a mean score of 3.52 (SD = 0.71) among the 45 participants with complete item-level responses. For interpretability, the mean score was linearly transformed from the 1–5 response scale to a 0–100 scale, yielding an index value of approximately 63.0. This value is presented as a transformed usability index rather than a conventional System Usability Scale (SUS) score because the instrument comprised 15 SUS-informed items rather than the standard 10-item SUS. Internal consistency of the scale was excellent (Cronbach's α = .92). Overall, participants generally perceived the simulated ECG-authentication prototype as learnable, usable, and suitable for integration into mobile applications.

## 6.2.2 Exploratory factor analysis

We conducted exploratory factor analysis using the 45 participants who provided complete responses to all 15 usability items. We reverse-scored negatively worded items before analysis so higher values consistently represented more favourable usability perceptions. The Kaiser–Meyer–Olkin measure indicated good sampling adequacy (KMO = .86), and Bartlett’s test of sphericity was statistically significant ($\chi^2(105) = 460.03$, $p < .001$), indicating that the correlation matrix was suitable for factor analysis.

We conducted exploratory factor analysis using minimum residual extraction and oblique rotation. Because usability burden and ease/integration were expected to be related constructs, an oblique rotation was selected. Parallel analysis supported retaining two factors. Together, the two-factor solution accounted for 56.24% of the rotated variance. The first factor accounted for 41.6% and comprised Q12, Q13, Q14, Q16, Q17, Q19, Q20, Q21, Q23, and Q25. These items primarily reflected perceived time, effort, complexity, inconsistency, technical support requirements, and learning burden, and were labeled Usability Burden. Because these items were negatively worded, we reverse-scored them before constructing the scale.

The second factor accounted for 14.7% and comprised Q15, Q18, Q22, Q24, and Q26. These items reflected intention to use, ease of use, learnability, confidence, and integration with mobile applications, and were labeled as Ease and Integration. All items had primary factor loadings of at least .53, providing a relatively clear two-factor interpretation. Given the small complete-case sample, however, treat the factor structure as exploratory and confirm it in a larger independent sample.

### 6.2.3 Reliability analysis

We assessed internal consistency using data from the 45 participants who provided complete responses to all 15 usability items. The results are depicted in Table 13. The full ECG usability scale demonstrated excellent internal consistency (Cronbach's α = .92). The 10-item Usability Burden factor also demonstrated excellent internal consistency (α = .94). In contrast, the five-item Ease and Integration factor demonstrated good internal consistency (α = .80). These findings should be interpreted as preliminary because the factor analysis was conducted using only 45 complete cases. Although the overall scale demonstrated excellent internal consistency (Cronbach's α = .92), replication in a larger independent sample is required before drawing conclusions regarding the stability of the factor structure or the reliability of the exploratory subscales.

*Table 13.* Reliability statistics for the ECG usability scale and exploratory subscales (N = 45 complete cases).

| Scale | Number of Items | Cronbach's α |
|---|---|---|
| **ECG usability scale** | 15 | .92 |
| **Usability burden factor** | 10 | .94 |
| **Ease and integration factor** | 5 | .80 |

### 6.2.4 Perceived security

Approximately 80% of participants perceived ECG authentication as more secure than traditional password-based authentication. When comparing ECG authentication with password-based mobile-device unlocking, 17 participants (34.0%) rated ECG authentication as much more secure, and 23 participants (46.0%) rated it as somewhat more secure. Only three participants (6.0%) considered ECG authentication less secure, while five participants (10.0%) perceived no difference and two participants (4.0%) reported that they did not know. Similarly, when participants compared ECG authentication with the passwords they typically use in their daily lives, 17 participants (34.0%) rated ECG authentication as much more secure, and 23 participants (46.0%) rated it as somewhat more secure. One participant (2.0%) considered ECG authentication somewhat less secure, one participant (2.0%) considered it much less secure, five participants (10.0%) perceived no difference, and three participants (6.0%) reported that they did not know. Overall, participants generally perceived the simulated ECG-authentication approach as more secure than password-based authentication. However, these findings reflect perceived security rather than the measured technical security, accuracy, or attack resistance of an operational ECG-authentication system.

### 6.2.5 Experience effects

Participants who reported using biometric authentication in their daily lives reported slightly higher usability scores than those who did not. Biometric users reported a mean usability score of **3.67** (SD = **0.75**, n = 17), whereas non-biometric users reported a mean score of **3.42** (SD = **0.68**, n = 28). This difference was not statistically significant, Welch's $t(31.16)$ = **1.13**, $p$ = **.269**. Although biometric users reported numerically higher usability scores, there was insufficient evidence that prior biometric experience was associated with significantly more favourable usability perceptions following prototype exposure.

### 6.2.6 Age-based analysis

To explore potential age-related differences in usability perceptions, we compared overall usability scores between participants younger than 55 years and those aged 55 years and older using the complete-case dataset (N = 45). Participants younger than 55 years reported a mean usability score of 3.46 (SD = 0.72, n = 40), whereas participants aged 55 years and older reported a mean usability score of 4.01 (SD = 0.32, n

= 5). This difference was statistically significant, Welch's $t(10.39) = -3.06$, $p = .012$. However, this finding should be interpreted as exploratory because the older-adult subgroup contained only five complete cases and was highly uneven relative to the younger group. Replication with a larger, more balanced sample is required before drawing conclusions about age-related usability differences.

### 6.2.6 Usability interpretation

The strongest positive usability pattern was that participants viewed the simulated ECG-authentication prototype as easy to use, highly learnable, potentially more convenient than password-based login, and well integrated with mobile applications. Participants also indicated a willingness to use ECG authentication in place of traditional passwords, provided that the technology was reliable, understandable, and practical to use. At the same time, survey responses suggested that acceptance was conditional rather than automatic, with participants continuing to consider issues such as usability, efficiency, and trust when evaluating the technology. These findings suggest that participants perceived the simulated ECG-authentication interaction favourably under controlled conditions. Further evaluation using an operational ECG-authentication system is required before conclusions can be drawn regarding real-world usability or viability. Successful adoption will likely depend on maintaining user confidence, clearly explaining the technology, preserving privacy, and offering appropriate fallback authentication methods.

The present research investigated public perceptions of ECG-based biometric authentication and participants' responses to a simulated ECG-authentication interface using a two-stage evaluation approach. Overall, the findings suggest that familiarity with biometric technologies was associated with greater willingness to adopt ECG authentication. Participants who interacted with the simulated prototype generally reported favourable perceptions of usability and perceived security under controlled conditions. Four key findings emerged from the study. First, password-related burden remains a persistent challenge for many users. Second, familiarity with biometric technologies was associated with greater acceptance of ECG authentication. Third, acceptance of ECG authentication was moderate and conditional, with participants also expressing concerns regarding privacy, transparency, reliability, and data governance. Fourth, participants who interacted with the simulated ECG-authentication prototype generally perceived the interaction as learnable, easy to use, and suitable for integration with mobile applications.

# 7. Discussion

This research examined public perceptions and usability of ECG-based biometric authentication through a two-stage pilot evaluation. Across the survey and prototype components, the findings point to a central theme: users are open to ECG authentication when it appears secure, learnable, and familiar, but acceptance depends on whether the system is understandable, trustworthy, and practical in everyday use. Because Studies 1 and 2 involved separate participant samples, the findings should be interpreted as complementary rather than as evidence that prototype interaction improved participants' perceptions.

The findings from Studies 1 and 2 reveal several interesting patterns. The first pattern suggests that password-related burdens remain a key challenge for users. A substantial proportion of participants reported password reuse and difficulty managing multiple credentials. This finding reinforces the need to evaluate passwordless authentication not only for technical security, but also for its ability to reduce everyday authentication burden.

Second, familiarity with biometric technologies appears to influence acceptance of ECG authentication. Participants with previous biometric experience tended to report more favourable perceptions of ECG authentication in Study 1 than those without such experience. Although not every comparison reached conventional statistical significance, the overall pattern suggests that acceptance of a novel biometric method may be shaped by prior exposure to more familiar biometric systems, such as fingerprint or facial recognition. In Study 2, participants who interacted with the prototype generally reported favourable perceptions of ECG authentication, particularly regarding ease of use, learnability, confidence, and integration with mobile applications. Together, the findings from the two studies suggest that both prior biometric familiarity and direct interaction with an ECG authentication prototype may shape how users evaluate this emerging technology.

An important pattern emerging across the two studies concerned the relationship between age, familiarity, and ECG authentication. In Study 1, age was not significantly associated with acceptance of ECG authentication, whereas prior biometric experience was associated with greater willingness to adopt ECG authentication. In Study 2, the small group of participants aged 55 years and older reported higher usability scores than younger participants. However, because only five older participants had complete usability-scale data, this result should be considered exploratory. The findings therefore caution against assuming that older adults are inherently resistant to passwordless authentication technologies, while also highlighting the need for larger and more balanced age-group comparisons.

A further pattern suggests that participants viewed ECG authentication positively, although not without critical reflection. Just over half of the participants in Study 1 indicated that they would be comfortable using ECG authentication. The remaining reluctance was not simply resistance to new technology. Participants raised specific concerns about how ECG authentication works, whether it would be reliable in practice, whether it would introduce additional complexity, and how biometric information would be stored and protected. Some participants also expressed concerns about the possibility of biometric information being hacked or misused. These findings suggest that acceptance of ECG authentication is shaped not only by perceived benefits, but also by concerns regarding trust, privacy, transparency, and security.

Participants in Study 2 generally reported favourable perceptions after interacting with the ECG authentication prototype. Many participants said the system was easy to learn and use, and relatively few believed they would need substantial technical support to use ECG authentication. Participants also responded positively to integrating ECG authentication into mobile applications, suggesting that ECG-based authentication may be more acceptable when implemented within familiar technologies and everyday digital experiences. Most participants felt confident using the system and did not believe it would require extensive training. Together, these findings provide preliminary evidence that users may view ECG authentication as learnable and practical following prototype-based interaction.

The usability evaluation provided additional evidence that ECG authentication is perceived as learnable and practical. Participants reported the strongest agreement for ease of use (Q18: $M = 3.94$, $SD = 0.85$), learnability (Q22: $M = 3.94$, $SD = 0.86$), and confidence in using ECG authentication (Q24: $M = 3.90$, $SD = 0.90$). Participants also responded positively to integration with mobile applications (Q26: $M = 3.82$, $SD = 0.67$). Ease of use and learnability received the highest mean ratings, followed by confidence and integration with mobile applications. In contrast, technical support requirements, perceived complexity, and efficiency-related concerns received the lowest ratings. Taken together, these findings suggest that implementation challenges, rather than fundamental usability limitations, may be the primary barriers to adoption.

Perceived security ratings were particularly strong. Approximately 80% of participants perceived ECG authentication as more secure than traditional password-based authentication. This finding is noteworthy because perceived security ratings were substantially stronger than overall usability ratings, suggesting that users may be willing to tolerate some additional effort or complexity when they perceive meaningful security benefits.

The open-text responses collected in Study 1 provided further insight into participants' concerns. Several respondents questioned whether ECG signals would remain sufficiently stable over time, particularly in the presence of illness, stress, arrhythmias, exercise, or other health-related changes. Others raised concerns about the sensitivity of ECG data and the possibility that health-related information could be exposed or misused. Participants interested in ECG authentication generally preferred implementations integrated into familiar devices, such as smartphones or smartwatches, rather than unfamiliar standalone hardware. Together, these comments suggest that long-term acceptance of ECG authentication will depend not only on usability and perceived security, but also on confidence in signal reliability, privacy protection, and practical deployment.

From a design perspective, the findings suggest that future ECG authentication systems should prioritise integration with familiar mobile ecosystems, minimise setup and support requirements, provide transparent explanations of data-handling practices, and include appropriate fallback authentication options. Given concerns about signal variability and health-related data, future deployments should also emphasise privacy-by-design principles and clearly communicate how biometric information is stored, protected, and used.

## 7.1 Implications for design

Future ECG authentication systems should prioritise integration into familiar mobile ecosystems, such as smartphones and smartwatches, rather than rely on standalone or implantable hardware. Participants generally responded positively to the idea of ECG authentication when it could be incorporated into devices and workflows they already use. Design efforts should minimise setup complexity and technical support requirements, as these were among the weakest usability dimensions identified in the prototype evaluation. Learnability emerged as one of the strongest usability characteristics, suggesting users generally viewed ECG authentication as approachable and relatively easy to understand after interacting with the prototype.

Study 1 identified prior biometric familiarity as an important factor associated with acceptance of ECG authentication. Participants who were already familiar with biometric technologies were more likely to report willingness to adopt ECG authentication. Consequently, onboarding processes, guided enrolment procedures, and educational materials that increase familiarity with biometric workflows may support wider adoption. Although participants with prior biometric experience reported numerically higher usability scores in Study 2, this difference was not statistically significant and should therefore be interpreted cautiously.

The qualitative responses also indicate that users require reassurance regarding signal reliability, particularly under conditions such as illness, stress, exercise, ageing, or cardiac irregularities. Several participants expressed concerns about whether ECG signals would remain sufficiently stable over time and whether health conditions might affect authentication performance. Transparent communication about how ECG templates are generated, stored, protected, and updated should therefore form a central component of system design.

Future ECG authentication systems should also include clear fallback authentication mechanisms and recovery options when ECG authentication cannot be completed successfully. Providing users with alternative authentication pathways may reduce reliability concerns while maintaining accessibility and inclusiveness.

Finally, because Study 1 found no significant age-related difference in ECG authentication acceptance, and the exploratory Study 2 usability comparison was based on only five older participants with complete usability-scale responses, future design research should prioritise larger, more balanced age-group samples. Current evidence suggests that familiarity, transparency, onboarding, and accessibility may matter more than chronological age alone. Designers should therefore focus on creating authentication systems that are understandable, learnable, and usable across diverse user populations rather than assuming age-related resistance to passwordless technologies.

## 7.2 Implications for cybersecurity practice

The findings also suggest that user education and onboarding may be important cybersecurity interventions in their own right. Participants with greater familiarity with biometric technologies generally reported more favourable attitudes toward ECG authentication, indicating that increasing user understanding may be as important as technical implementation. Particular attention should be given to supporting users whose physiological conditions may affect ECG consistency over time. Accessible recovery mechanisms, alternative authentication pathways, and clear error-handling procedures may be especially important when illness, stress, exercise, aging, or cardiac conditions influence ECG patterns.

## 7.3 Implications for policy and health-data governance

The findings highlight the importance of treating ECG authentication data as sensitive personal information. Participants frequently expressed concerns regarding privacy, data storage, and the possibility that ECG data could be compromised or linked to health information. Consequently, future deployments should adopt privacy-by-design principles, including transparent consent procedures, strong template protection, secure storage practices, and clearly defined breach-response frameworks. Where possible, ECG templates should be stored locally or in pseudonymised form to minimise privacy risks.

The qualitative responses further suggest that users perceive ECG data as fundamentally different from other biometric identifiers because of its association with health and medical information. Several participants expressed concerns regarding the potential misuse of health-related information. They questioned whether ECG authentication data could be linked to clinical records or used to infer aspects of an individual's health status. Policymakers and organisations should therefore ensure that authentication data remain clearly separated from health records and that users are informed about how ECG data are collected, processed, stored, retained, and protected.

The findings also highlight the importance of transparency and user control. Many participants reported uncertainty regarding how ECG authentication works and how biometric information would be managed. Clear explanations of data collection, storage, retention, deletion, and access rights may therefore be essential to building trust in ECG-based authentication systems.

Because Study 1 found no significant age-related difference in ECG acceptance, and the Study 2 usability comparison was based on a very small older-adult subgroup, policy frameworks should avoid assuming that chronological age is itself a barrier to adoption. Instead, accessibility, inclusion, user education, and opportunities to build familiarity should remain central considerations.

## 7.4 Comparison with existing biometric authentication literature

The pre-prototype ECG acceptance rate of 52.2% suggests moderate willingness to adopt ECG authentication. In Study 1, prior biometric experience and perceived biometric security were associated with ECG acceptance. In Study 2, participants with prior biometric experience reported numerically higher usability scores, although this difference was not statistically significant. Approximately 80% of participants in the usability study perceived ECG authentication as more secure than traditional password-based authentication. These findings align with previous biometric-authentication research showing that familiarity, perceived convenience, and perceived security influence adoption attitudes (Kružíková et al., 2022; Wolf et al., 2019; Wang et al., 2024).

However, the present study also highlights several considerations that appear particularly relevant to ECG authentication. Unlike more established biometric approaches such as fingerprint or facial recognition, ECG authentication is closely associated with health-related information. Participants frequently expressed concerns about the stability of ECG signals over time, the impact of illness or cardiac conditions on authentication, and the potential for ECG data to reveal sensitive health information. These concerns suggest that ECG authentication occupies a distinct position of trust within the broader biometric authentication landscape, where perceptions of privacy, transparency, and health data governance may matter most.

The findings also extend previous research on authentication and aging. In Study 1, older adults reported lower levels of biometric adoption than younger participants, but age was not significantly associated with ECG acceptance. In Study 2, the older subgroup reported higher usability scores, although this exploratory comparison included only five older complete cases. Larger, more balanced evaluations are therefore needed to determine whether age influences ECG-authentication usability in broader populations.

## 7.5 Older adults, familiarity, and inclusive authentication design

The age-related findings provide preliminary insight into the factors influencing ECG-authentication acceptance and usability. In Study 1, older adults reported lower levels of prior biometric adoption than younger adults, yet they were not significantly less willing to adopt ECG authentication. In Study 2, participants aged 55 years and older reported higher mean usability scores than younger participants. However, only five older participants provided complete usability-scale responses, making this result sensitive to individual observations and unsuitable for broad generalisation.

Collectively, the findings do not support the assumption that older adults are inherently resistant to passwordless authentication technologies. They instead highlight the need for larger age-balanced usability studies and for authentication systems that are transparent, learnable, accessible, and supported by appropriate onboarding and recovery mechanisms. Familiarity and exposure remain relevant considerations, but their relationship with usability requires further testing in a larger Study 2 sample.

## 7.6 Privacy, transparency, and user control

One of the most important usable security findings is that participants' concerns focused less on authentication effectiveness and more on transparency, privacy, and control over biometric information. The most frequently reported barriers were lack of understanding, uncertainty about how ECG authentication works, perceived complexity, discomfort with storing biometric information, and concern that biometric templates could be compromised. The available open-text responses reinforced these concerns, with several participants describing ECG data as medically sensitive and expressing uncertainty about how such information would be stored, protected, and potentially linked to other personal records. Far fewer comments focused on authentication failure or signal accuracy, suggesting that trust and privacy concerns may be more influential than technical performance in shaping user perceptions.

These findings suggest that successful deployment of ECG authentication will require more than technical accuracy. Users need clear, plain-language explanations of what ECG data are collected, how biometric templates are stored and protected, who can access them, whether any health-related information could potentially be inferred, and what options are available if ECG authentication cannot be used. Transparent consent processes, clear privacy protections, and accessible fallback authentication methods may therefore be critical to building user confidence and supporting wider adoption.

# 8. Limitations

This study has several limitations. First, Study 1 used a convenience sample and cross-sectional survey design, which limits causal interpretation and broader generalisability. Second, the survey and usability samples were not directly linked; therefore, the manuscript does not claim that the same individuals moved from pre-prototype attitudes to post-prototype usability perceptions. Third, Study 2 was exploratory and involved 49 completed post-test responses, of which 45 contained complete data across all 15 usability items. Representation across age groups was uneven, and only five participants aged 55 years and older were included in the complete-case usability analysis. Consequently, the Study 2 findings, particularly the age comparison, should be interpreted as preliminary rather than definitive evidence regarding broader populations. Fourth, the ECG process was evaluated through a prototype rather than a fully deployed ECG-enabled system, meaning that the findings reflect perceived usability, perceived security, and user attitudes rather than real-world authentication performance, long-term usability, signal stability, or resistance to attack. Fifth, the exploratory factor analysis used 45 complete cases across 15 usability items. Although sampling-adequacy statistics were favourable and the two-factor loading pattern was interpretable, the small participant-to-item ratio limits the stability and generalisability of the factor solution. The proposed factor structure should therefore be confirmed in a larger independent sample. Finally, participants were recruited through an open call, which may have introduced self-selection bias towards individuals who were more comfortable with digital technologies or more interested in cybersecurity and authentication.

# 9. Conclusion

Taken together, the findings provide moderate but conditional support for ECG-based biometric authentication. Participants recognised the limitations of password-based authentication and generally perceived biometrics as more secure. Yet, their willingness to adopt ECG authentication depended on familiarity, perceived reliability, privacy protection, and practical implementation considerations. Across both studies, familiarity with biometric technologies emerged as a more consistent influence on acceptance than chronological age, suggesting that exposure, education, and onboarding may help support adoption. Participants who interacted with the prototype generally reported positive perceptions of usability and security, suggesting that ECG authentication may be viewed as a learnable, usable alternative to traditional password-based authentication when implemented through familiar devices and workflows. At the same time, concerns regarding transparency, privacy, health-related data, and signal reliability underscore the importance of user-centred design and robust governance frameworks. Overall, the simulated ECG-authentication concept was generally perceived positively by participants, although further evaluation using operational ECG-authentication systems is required before conclusions can be drawn regarding practical deployment or viability.

# 10. Future Research

Future research will extend this work through longitudinal studies, real-world ECG-enabled deployments, larger and more demographically balanced usability evaluations, and direct comparisons with widely used biometric methods such as Face ID and fingerprint authentication. We also aim to investigate privacy-preserving ECG template storage, failure recovery, accessibility for older adults and users with health conditions, and whether the logistic regression findings replicate in larger, more diverse samples. In particular, we aim to examine whether prior biometric use remains an independent predictor of ECG acceptance in real-world deployments. A further priority is to develop and test a full extended TAM model for ECG authentication using validated multi-item scales for perceived usefulness, perceived ease of use, perceived security, trust, perceived risk, privacy concern, and behavioural intention to use.

# Declarations

## Ethics Statement

The Western Sydney University Human Research Ethics Committee approved this study (Approval No. H15201). All participants were aged 18 years or older and provided informed consent before participating. The study followed the University's human research ethics requirements and relevant ethical guidelines.

## Funding

This research was supported by a Western Sydney University Early Career Researcher (ECR) Fellowship Grant.

## Acknowledgements

The authors sincerely thank all participants for their time, engagement, and valuable contributions to this study. Their willingness to share their experiences and perspectives made this research possible.

## Data Availability

The data underlying this study are not publicly available because of ethics and participant privacy considerations. De-identified data may be available from the corresponding author upon reasonable request and subject to approval consistent with the conditions of ethics approval H15201.

## Conflict of Interest

The authors declare no conflicts of interest.

## Author Contributions

Farnaz Farid: Conceptualisation, methodology, investigation, formal analysis, project administration, writing—original draft, writing—review and editing.
Linet Momposhi: Investigation, data collection, writing—review and editing.
Tamara Watson: Conceptualisation, supervision, methodology, writing—review and editing.
Alana Maurushat: Supervision, methodology, writing—review and editing.
Abubakar Bello: Methodology, formal analysis, writing—review and editing.
Farhad Ahamed: Writing—review and editing.
All authors read and approved the final manuscript.